# Dictionary for Sparse Representation of Chirp Echo in Broadband Radar

Lei GAO[1]

ATR Lab,NUDT;Changsha,Hunan,410072, P.R.China.

**Abstract**: A new dictionary $D$ for sparse representation of chirp echo in broadband radar is put forward in this paper. Different from chirplet decomposition which decomposes echo in time-frequency plane, the dictionary $D$ transforms the sparsity of target observed by radar in distance range to the sparsity in frequency domain by stretch processing and the sparse representation of echo is realized. With strict mathematical deduction, the sparsity of echo in dictionary $D$ is proved and the dictionary is orthogonal. As to the property of dictionary for application, its construction is simple, the necessary parameters for it can be obtained conveniently and it is convenient for using. Furthermore, the object of application can be extended to the echo of multi-component chirps with single freedom degree.

## I. Introduction

Chirp signal is common style in broadband radar, for example, in Inverse Synthetical Aperture radar. Considering following signal model:

$$s(t) = \exp\left[ j2\pi\left( f_c t + \gamma t^2/2 \right) \right], t \in [-T/2, T/2] \qquad (1)$$

In which, $f_c$ is carrier frequency, $T$ is pulse width, $\gamma$ is the rate of frequency modulation. The bandwidth is $B = \gamma T$. In order to ensure certain resolution in range, the bandwidth can reach several hundred mega-hertz or thousand mega-hertz[1]. There are two kinds of methods for processing this broadband signal, one is to process the signal in analog domain using stretch processing[2], and the other is to sample directly using the theorem of bandpass sampling[3]. According with the theorem of bandpass sampling, the sampling frequency $f_s \geq 2B$, and the length of discrete data from sampling is $f_s T$. For example, the bandwidth of nowadays can be $B = 1\text{GHz}$, the pulse width is $T = 50\mu s$, then the length of discrete data isn't less than 10000. Using long integer with 32 bits to denote one number, then the data volume for one piece of pulse is about $40\text{KB}$. If the pulse repetition frequency is $1000\text{Hz}$, then there are 500 frames of echo pulse in one second(generally, the broadband and narrowband pulse are emitted alternately in the system of broadband radar, the narrowband pulse is used for tracking), the data volume is about $20\text{MB}$ and the data volume of echo is $72\text{GB}$ when the radar system works continuously for an hour. It is an enormous challenge to data storage.

From the compressed sensing theory put forward by Candes,Romberg,Tao[4][5] and Donoho[6], it can be known that for sparse signal in some sparsity basis, it can be compressed with random projection and reconstructed with high probability. If the echo is sparse under some dictionary (basis), then it can be compressed using compressed sensing theory to reduce the space for storage and can be reconstructed with the theory when it needs to process.

The foregoing literatures considered the sparsity of chirp signal from the view of chirplet decomposition, but the dictionary constructed by chirplet decomposition has high redundancy[7] and the process for construction is complicated. Although from the latest research result by Candes[8], it can be known that the dictionary can be redundant, the excessively hugeness of the dictionary constructed by chirplet decomposition and the complexity of constructing process cause that the dictionary is not convenient for applying to process of echo of broadband radar. In this paper, a new orthogonal dictionary is put forward for the sparse representation of echo. Different from chirplet decomposition which decomposed echo in time-frequency plane and obtained the sparse representation of echo[9], the new dictionary transforms the sparsity of target observed by radar in distance range to the sparsity in frequency domain by stretch processing and the sparse representation of echo can be obtained. The new dictionary has clear physical meanings, simple structure and better practicability.

---

[1] Email:ren_lgao@126.com

The paper is organized as follows. In section two, the detailed construction of new dictionary is given and the sparsity of echo on the dictionary is proved in mathematics. In section three, the inherent sparsity of echo is analyzed from the course of observation for broadband radar and connection between sparsity and the dictionary is explained. In the section four, the practicability and the extensibility of object for application are discussed. In the end, it is the summarization.

## II. Construction of Dictionary

Stretch processing is common method to process the echo in analog domain. Consider that the target accords with point scatterer model, under the radiation of signal denoted by Eq.(1), then the echo can be written as:

$$s_r(t) = \sum_{i=1}^{P} A_i \exp\left\{ j2\pi \left[ f_c(t-t_{di}) + \frac{1}{2}\gamma(t-t_{di})^2 \right] \right\} \quad (2)$$

In which, $P$ is the number of scatterers, $A_i, t_{di}$ denote the scattering intension and delay for the $i$th scatterer respectively. $t_{di} = 2r_i/c$, $r_i$ is the distance of the $i$th scatterer to radar and $c$ is the velocity of electromagnetic wave in vacuum.

The reference signal for stretch processing is[2]:

$$s_{ref}(t) = \exp\left\{ j2\pi \left[ f_c(t-t_{ref}) + \frac{1}{2}\gamma(t-t_{ref})^2 \right] \right\} \quad (3)$$

The result of stretch processing is intermediate frequency (IF) signal: $s_{IF}(t) = s_r(t) \times [s_{ref}(t)]^*$, $[\ ]^*$ means conjugate.

From deduction $s_{IF}(t)$ can be written as:

$$s_{IF}(t) = \sum_{i=1}^{P} A_i \exp\left\{ -j2\pi \left[ f_c(t_{di}-t_{ref}) + \gamma t(t_{di}-t_{ref}) - \frac{1}{2}\gamma(t_{di}^2 - t_{ref}^2) \right] \right\} \quad (4)$$

In Eq. (2)(3)(4), $t \in [-T/2, T/2]$. From Eq. (4), $s_{IF}(t)$ is constitutive of several single frequency signals and sparse in frequency domain. The number of frequencies in $s_{IF}(t)$ is $P$.

Sampling $t \in [-T/2, T/2]$, $\mathbf{t} = [-T/2, -T/2+1/f_s, -T/2+2/f_s, ..., -T/2+(N-1)/f_s]^T = [\mathbf{t}(1), \mathbf{t}(2), \mathbf{t}(3), ..., \mathbf{t}(N)]^T$ can be obtained. Of which $(N-1)/f_s \leq T < N/f_s$. The sampling result of $s_r(t)$, $s_{ref}(t)$ and $s_{IF}(t)$ can be written in vector form:

$$\mathbf{s\_r} = \left\{ \left[ \sum_{i=1}^{P} A_i \exp\left\{ j2\pi \left[ f_c(\mathbf{t}(n)-t_{di}) + \frac{1}{2}\gamma(\mathbf{t}(n)-t_{di})^2 \right] \right\} \right]_{1 \leq n \leq N, n \in \mathbb{N}} \right\}^T$$

$$\mathbf{s\_ref} = \left\{ \left[ \exp\left\{ j2\pi \left[ f_c(\mathbf{t}(n)-t_{ref}) + \frac{1}{2}\gamma(\mathbf{t}(n)-t_{ref})^2 \right] \right\} \right]_{1 \leq n \leq N, n \in \mathbb{N}} \right\}^T$$

$$\mathbf{s\_IF} = \left\{ \left[ \sum_{i=1}^{P} A_i \exp\left\{ j2\pi \left[ f_c(t_{di}-t_{ref}) + \gamma \times \mathbf{t}(n) \times (t_{di}-t_{ref}) - \frac{1}{2}\gamma(t_{di}^2 - t_{ref}^2) \right] \right\} \right]_{1 \leq n \leq N, n \in \mathbb{N}} \right\}^T$$

The dictionary $\mathbf{D}$ for sparse representation of $\mathbf{s\_r}$ can be obtained by following two steps.

The first step, construct $N \times N$ diagonal matrix from $\mathbf{s\_ref}$, the cells of the matrix satisfy:

$$\mathbf{\Phi}(m,n) = \begin{cases} \mathbf{s\_ref}(m), m = n \\ 0, m \neq n \end{cases} \quad (5)$$

In which, $m, n$ are the indexes of cell in matrix respectively and $1 \leq m, n \leq N, m, n \in \mathbb{N}$.

The second step, calculate the dictionary for signal that is sparse in frequency domain with Fast Fourier Transform (FFT) [10]:

$$\Psi = \frac{1}{\sqrt{N}}[\psi_0, \psi_1, \ldots \psi_{N-1}] \quad (6)$$

In which, $\psi_n = \left[\varphi_0^{(n)}, \varphi_1^{(n)}, \ldots, \varphi_{N-1}^{(n)}\right]^T$, $\varphi_m^{(n)} = \exp\left\{-j2\pi\frac{m}{N}n\right\}$, $1 \leq m, n \leq N, m, n \in \mathbb{N}$.

Then the dictionary is $\mathbf{D} = \mathbf{\Phi\Psi}$. In following context, it will be proved that the matrix $\mathbf{D} = \mathbf{\Phi\Psi}$ can be acted as the dictionary for the sparse representation of echo. That is to prove the following proposition.

***Proposition I: The vector* s_r *is sparse in the matrix* $\mathbf{\Phi\Psi}$.**

**Proof**: From the stretch processing, the following equation can be obtained:

$$\mathbf{s\_IF} = \mathbf{\Phi}^H \times \mathbf{s\_r} \quad (7)$$

From the expression of **s_IF**, it is constituted by several single frequency signal. The analog frequencies are $\gamma(t_{di} - t_{ref}), i = 1, \ldots, P$ and the corresponding digital frequencies are $2\pi \times \gamma(t_{di} - t_{ref})/f_s, i = 1, \ldots, P$. $P$ is finite number, so **s_IF** is sparse in dictionary $\Psi$. On the assumption that the sparse vector of **s_IF** in dictionary $\Psi$ is $\boldsymbol{\alpha}$, then $\mathbf{s\_IF} = \mathbf{\Psi\alpha}$. Combine with Eq. (7), the following equation can be obtained:

$$\mathbf{s\_IF} = \mathbf{\Phi}^H \times \mathbf{s\_r} = \mathbf{\Psi\alpha} \quad (8)$$

From the construction of $\mathbf{\Phi}$, it can be proved easily that $\mathbf{\Phi}^H \mathbf{\Phi} = \mathbf{I}$ is true, in which $\mathbf{I}$ is $N \times N$ identity matrix. Then, $\mathbf{\Phi}$ is invertible and $\mathbf{\Phi}^{-1} = \mathbf{\Phi}^H$. From Eq. (8), it can be deduced that:

$$\mathbf{s\_r} = \mathbf{\Phi\Psi\alpha} \quad (9)$$

The $\boldsymbol{\alpha}$ is sparse vector. So **s_r** is sparse in the matrix $\mathbf{\Phi\Psi}$. This completes the proof.

***Corollary I. The dictionary* $\mathbf{D}$ *is orthogonal.***

**Proof:** From the construction Eq.(6) for dictionary $\mathbf{\Psi}$ by FFT, it can be proved that $\mathbf{\Psi}^H \mathbf{\Psi} = \mathbf{I}$ is true, so,

$$\mathbf{D}^H \mathbf{D} = \left(\mathbf{\Psi}^H \mathbf{\Phi}^H\right)\mathbf{\Phi\Psi} = \mathbf{I} \quad (10)$$

Then the dictionary $\mathbf{D}$ is orthogonal. This completes the proof.

Because of the orthogonality of dictionary, from Eq. (9), the calculation formula for sparse coefficient is:

$$\boldsymbol{\alpha} = \mathbf{D}^H \mathbf{s\_r} = \mathbf{\Psi}^H \mathbf{\Phi}^H \mathbf{s\_r} \quad (11)$$

It means that it is convenient to obtain sparse coefficient and complex algorithm for sparse representation is needless. Furthermore, the orthogonality also lessen the restrict on the sensing matrix for compression when the dictionary is integrated into compressed sensing theory[6][11].

## III. Physical Interpretation of Sparsity for Echo in Broadband Radar

### A. Analysis of Sparsity for Echo in Broadband Radar

Generally, when the radar observes the target, in one period of time, the distance observed has certain range, for instance, the range is $[R_t, R_h]$, $R_t < R_h$. For single pulse, the length of distance range is no less than the distance of electromagnetic wave moved in the pulse width, it means $(R_h - R_t) \geq cT$. If the pulse width $T = 50\mu s$, then $(R_h - R_t) \geq 15\text{km}$. For several pulses, the length of distance range will be more. Divide the $[R_t, R_h]$ with the range resolution of radar: $\Delta R = c/(2B)$, the assemble of distance interval can be obtained as $\text{Ran}(R) = \{[R_i, R_{i+1})\}_{1 \leq i \leq M, i \in \mathbb{N}}$, in which $R_1 = R_t, R_{i+1} = R_i + \Delta R$, and $M$ is fixed by the inequality $M \times \Delta R \leq (R_h - R_t) < (M+1) \times \Delta R$. For each distance interval, using $R_i$ to denote the distance of all location in this distance interval to radar, then $\{t_i\}_{1 \leq i \leq M, i \in \mathbb{N}}$ denotes the assemble of delay for all possible scatterers in $[R_t, R_h]$, in which $t_i = 2R_i/c$.

Although the distance range that radar observed is $[R_t, R_h]$, its length can be several kilometers or more, the size of the target observed by radar is finite, for instance, the size of plane and man-made orb in space is tens of meters and no more than hundred meters. So the distance range in which the scatterers exist belongs to small number of distance interval in the assemble $\text{Ran}(R)$. If the number of scatterers is $P$, then the number of distance interval in which scatterers exist is $P'$ and $P' \leq P$, i.e. the number of distance interval in which scatterers exist is finite. This reflects the sparsity of the observed target in distance range.

Sparsity can be reflected in the expression of echo. On the assumption that the assemble of distance interval in which scatterers exist is $\{[R_l^{(s)}, R_l^{(s)} + \Delta R]\}_{l=1,\ldots,P', l \in \mathbb{N}}$ and $\{[R_l^{(s)}, R_l^{(s)} + \Delta R]\}_{l=1,\ldots,P', l \in \mathbb{N}} \subset \text{Ran}(R)$. The superscript of $R_l^{(s)}$ means that scatterers exist. The corresponding assemble of delay of echo is $\{t_l^{(s)}\}_{l=1,\ldots,P', l \in \mathbb{N}}$, then echo can be expressed as:

$$s_r(t) = \sum_{l=1}^{P'} A_l^{(s)} \cdot s\left(t - t_l^{(s)}\right) \tag{12}$$

In which $A_l^{(s)}$ is the intensity of scatterers in the $l$ th distance interval $[R_l^{(s)}, R_l^{(s)} + \Delta R]$. The last equation is the result of Eq.(2) after combining the echo of the scatterers with the same delay. It is knowable that the $\{t_l^{(s)}\}_{l=1,\ldots,P', l \in \mathbb{N}}$ is subclass of $\{t_i\}_{1 \leq i \leq M, i \in \mathbb{N}}$, so the echo expressed by (12) can be denoted by linear combination of $P'$ cells in the assemble $\{s(t - t_i)\}_{1 \leq i \leq M, i \in \mathbb{N}}$ with intensity of scatterers. Viz, the echo is sparse. The Eq.(12) is a kind of sparse representation if $\{s(t - t_i)\}_{1 \leq i \leq M, i \in \mathbb{N}}$ is considered as a sparsity basis.

**B Translation of Sparsity in Dictionary D**

Though the Eq. (12) is sparse representation of echo, it is the direct application of the sparsity of target observed by radar in the distance range and it is inconvenient for application, because under this circumstance the sparsity basis should be adjusted according to the distance range that radar observed.

The dictionary $\mathbf{D} = \mathbf{\Phi\Psi}$ makes use of the sparsity by translation. After simple analysis on Eq. (11), it can be found that evolvement from signal vector $\mathbf{s\_r}$ to sparse vector $\mathbf{\alpha}$ can be divided into two steps, first, $\mathbf{\Phi}^H$ transforms the signal $\mathbf{s\_r}$ that is sparse in distance range into signal $\mathbf{s\_IF}$ which is sparse in frequency domain, second, $\mathbf{\Psi}^H$ transforms signal $\mathbf{s\_IF}$ which is sparse in frequency domain into sparse vector. The translating process in Eq. (11) also shows the physical meanings of dictionary $\mathbf{D}$: transform the sparsity in distance range into sparse vector by two steps: stretch processing and FFT. It also reveals an idea that the sparse representation can be obtained by several steps of translation under certain sparsity.

## IV. Application Property of the Dictionary

Application property of the dictionary $\mathbf{D} = \mathbf{\Phi\Psi}$ mainly considers the practicability and the extensibility of object for application. The acquirement for dictionary in actual application is analyzed in the discussion of practicability. It can be found that the parameters of dictionary can be obtained conveniently and the dictionary $\mathbf{D}$ can be calculated directly if needed. In the discussion of the extensibility of object for application, the object for application is extended to the echo of multi-component chirps with single freedom degree. This signal form can be applied to radar which emits frequency-stepped chirps signal[12].

## A. Practicability

The practicability can be analyzed from the way of acquiring for dictionary. The dictionary **D** is constitutive of two parts: **Φ** and **Ψ** and the way of acquirement will be analyzed respectively.

The **Ψ** is only restricted by the dimension of signal vector. Generally, the dimension of signal vector for processing is fixed and the **Ψ** can be calculated by Eq.(6) when it is needed.

The **Φ** is restricted by two parts of parameters, one is parameters of signal emitted by radar, the other is the delay of reference signal $t_{ref}$. The parameters of signal emitted by radar are fixed, but $t_{ref}$ needs an appropriate range.

The discrete style of IF signal **s_IF** is obtained by multiplication of **s_r** and conjugate of **s_ref** which is sampled from Eq.(2) and Eq.(3) under the sampling frequency $f_s$ respectively. It is equivalent that the **s_IF** is obtained by sampling from $s_{IF}(t)$ with sampling frequency $f_s$. From Eq.(4), it can be known that the analog frequencies of $s_{IF}(t)$ are $\gamma(t_{di} - t_{ref}), i = 1, ..., P$. In order to avoid commixture of spectrum, the following inequation must be true[13]:

$$f_s \geq 2\gamma |t_{di} - t_{ref}|, i = 1, ..., P \tag{13}$$

Then the appropriate interval of $t_{ref}$ is $[\max_i(t_{di}) - f_s/(2\gamma), \min_i(t_{di}) + f_s/(2\gamma)]$ and selection of $t_{ref}$ must be according with the distance range of the target. The sampling frequency $f_s \geq 2B$, so the length of appropriate interval of $t_{ref}$ is $f_s/(B/T) - (\max_i(t_{di}) - \min_i(t_{di})) \geq 2T - (\max_i(t_{di}) - \min_i(t_{di}))$. The distance range of the target is small, and generally, $2T \gg (\max_i(t_{di}) - \min_i(t_{di}))$ is true. So, the selection of $t_{ref}$ has certain freedom space and the approximately distance range of target will satisfy the requirement. The approximate distance range of target can be obtained from narrowband tracking system which assists the broadband observation.

Integrate the above description, the two parts **Φ**, **Ψ** that constitute the dictionary **D** are not necessary for storage and they can be calculated directly if necessary. Beside that it needs to obtain approximately distance range of target from narrowband tracking system, the dictionary is restricted by the parameters of chirp signal and this property determines that the dictionary has better practicability.

## B Extensibility of Object for Application

From the ***Proposition I***, it can be known that the dictionary **D** can be applied to the echo of chirp signal. After analysis, it can be found that the object for application can be extended to the echo of multi-component chirps with single freedom degree

The multi-component chirps with single freedom degree are chirps among which the rate of frequency modulation, bandwidth and pulse width are all the same except carrier frequency. Consider chirps signal with the number of multi-tone is $K$, under the scatterer model above, the echo can be written as:

$$s_{r\_K}(t) = \sum_{k=1}^{K} \sum_{i=1}^{P} A_i \exp\left\{ j2\pi \left[ f_c^{(k)}(t - t_{di}) + \frac{1}{2}\gamma(t - t_{di})^2 \right] \right\} \tag{14}$$

In which $f_c^{(k)}, k = 1, ..., K$ are carrier frequencies. The other parameters are the same with above.

Still using the reference signal denoted by (3), then result of stretch processing from $s_{r\_K}(t)$ and $s_{ref}(t)$ can be written as:

$$s_{IF\_K}(t) = \sum_{k=1}^{K}\sum_{i=1}^{P} A_i \exp\left\{-j2\pi\left[\left(f_c^{(k)}t_{di} - f_c t_{ref}\right) + t\left(f_c - f_c^{(k)} + \gamma t_{di} - \gamma t_{ref}\right) - \frac{1}{2}\gamma\left(t_{di}^2 - t_{ref}^2\right)\right]\right\} \quad (15)$$

Analyze the last equation, $s_{IF\_K}(t)$ is constitutive of several single frequency signal still and its discrete style is sparse in $\Psi$. So the discrete style of $s_{r\_K}(t)$ is sparse in $\Phi\Psi$ and the dictionary $\mathbf{D}$ can be applied to sparse representation of multi-component chirps with single freedom degree.

## V. Conclusion

A new dictionary which can be applied to sparse representation of chirp echo in broadband radar is put forward. The dictionary has a simple structure and explicit physical meanings which has close connection with radar observation. In the application property, the needed parameters can be obtained conveniently, the construction of dictionary is simple and the dictionary is convenient to use. The object of application can be exptended to the echo of multi-component chirps with single freedom degree. The dictionary given in this paper can be combined with compressed sensing theory, and good compression may be realized to reduce the pressure for data storage caused by direct bandpass sampling of echo for broadband radar. It will be researched later.

Furthermore, sparse representation for echo of multi-component chirps with different carrier frequencies and different rate of frequency modulation can also be researched ulteriorly according to the idea of translation of sparsity which is used in the construction of dictionary in this paper.